\def\year{2020}\relax
\documentclass[letterpaper]{article} 
\usepackage{aaai20}  
\usepackage{times}  
\usepackage{helvet} 
\usepackage{courier}  
\usepackage[hyphens]{url}  
\usepackage{graphicx} 
\usepackage{graphicx}  
\usepackage{xcolor} 
\usepackage{amsmath} 
\usepackage{amssymb} 
\usepackage{soul} 
\usepackage{booktabs} 
\usepackage{subfig} 
\title{Hey Dona! Can you help me with student course registration?}
 \author{Vishesh Kalvakurthi\\
 Dept. of Computer Science\\
 Montclair State University \\
 Montclair, NJ, USA\\
 kalvakurthiv1@montclair.edu
 \And
 Aparna S. Varde\\
 Dept. of Computer Science, \\
 Faculty, Computational Linguistics\\
 Montclair State University, NJ, USA\\
 vardea@montclair.edu
 \And
 John Jenq\\
 Dept. of Computer Science\\
 Montclair State University\\ 
 Montclair, NJ, USA\\
 jenqj@montclair.edu
 }
\begin{document}
\newcommand{\ourmodel}[1]{\textsc{CSK-SNIFFER}}
\newcommand{\yolo}[1]{\textsc{YOLO}}
 
\definecolor{Red}{rgb}{1,0,0}
\definecolor{Green}{rgb}{0,0.7,0}
\definecolor{Blue}{rgb}{0,0,1}
\definecolor{Red}{rgb}{0.6,0,0}
\definecolor{Orange}{rgb}{1,0.5,0}
\newcommand{\niket}[1]{\textcolor{Green}{[#1 \textsc{--Niket}]}}
\newcommand{\anurag}[1]{\textcolor{Blue}{[#1 \textsc{--Anurag}]}}
\newcommand{\aparna}[1]{\textcolor{Red}{[#1 \textsc{--Aparna}]}}
\newcommand{\reviewed}[1]{\textcolor{Orange}{[#1]}}

\makeatletter
\newcommand*\bigcdot{\mathpalette\bigcdot@{.5}}
\newcommand*\bigcdot@[2]{\mathbin{\vcenter{\hbox{\scalebox{#2}{$\m@th#1\bullet$}}}}}
\makeatother

\renewcommand{\v}[1]{$\mathbf{#1}$}
\newcommand{\vect}[1]{\mathbf{#1}}

\newcommand{\bluebox}[1]{\colorbox{blue!10}{#1}}
\newcommand{\redbox}[1]{\colorbox{red!10}{#1}}
\newcommand{\purplebox}[1]{\colorbox{purple!10}{#1}}



\def\DG{{\mathcal{G}}}

\newtheorem{theorem}{Definition}[section]

\newcommand{\statechange}[1]{\texttt{\textit{#1}}}
\newcommand{\entity}[1]{\texttt{#1}}
\newcommand{\strikethrough}[1]{\st{#1}}

\newcommand{\namecite}[1]{\citeauthor{#1}~\shortcite{#1}}
\newcommand{\com}[1]{}
\newcommand{\myparagraph}[1]{\vspace{1mm} \noindent {\bf #1: }}
\newcommand{\bfit}[1]{\textbf{\textit{#1}}}
\newcommand{\eat}[1]{}
\mathchardef\mhyphen="2D
\newenvironment{ite}{                     
     \parskip 0cm \begin{itemize} \parskip 0cm \parsep 0cm \itemsep 0cm \topsep 0cm}{
        \end{itemize}} 
\newenvironment{enu}{                   
     \parskip 0cm \begin{list}{}{\parsep 0cm \itemsep 0cm \topsep 0cm}}{
      \end{list}} 
\newenvironment{des}{                 
     \parskip 0cm \begin{list}{}{\parsep 0cm \itemsep 0cm \topsep 0cm}}{
      \end{list}} 
\newenvironment{myenumerate}{                   
     \parskip 0cm \begin{enumerate}{\parsep 0cm \itemsep 0cm \topsep 0cm}}{
        \end{enumerate}} 
\newenvironment{myitemize}{                     
     \parskip 0cm \begin{itemize}{\parsep 0cm \itemsep 0cm \topsep 0cm}}{
        \end{itemize}} 
\newcommand{\ourdataexpansion}{``What-If Question Answering''}
\newenvironment{myquote}{                   
  \parskip 0mm \begin{quoting}[vskip=0mm,leftmargin=2mm]}{
\end{quoting}}
\newcommand{\red}[1]{\textcolor{red}{#1}}
\newcommand{\blue}[1]{\textcolor{blue}{#1}}
\newcommand{\green}[1]{\textcolor{green}{#1}}
\newenvironment{mycentering}
 {\parskip=0pt\par\nopagebreak\centering}
 {\par\noindent\ignorespacesafterend}

    

\maketitle
\begin{abstract}
In this paper. we present a demo of an intelligent personal agent \emph{Hey Dona} or just \emph{Dona} with virtual voice assistance in student course registration. It is a deployed project in the theme of AI for education. In this digital age with a myriad of smart devices, users often delegate tasks to agents. While pointing and clicking supersedes the erstwhile command-typing, modern devices allow users to speak commands for agents to execute tasks, enhancing speed and convenience. In line with this progress, Dona is an intelligent agent catering to student needs by automated, voice-operated course registration, spanning a multitude of accents, entailing task planning optimization, with some language translation as needed. Dona accepts voice input by microphone (Bluetooth, wired microphone), converts human voice to computer understandable language, performs query processing as per user commands, connects with the Web to search for answers, models task dependencies, imbibes quality control, and conveys output by speaking to users as well as displaying text, thus enabling human-AI interaction by speech cum text. It is meant to work seamlessly on desktops, smartphones etc. and in indoor / outdoor settings. To the best of our knowledge, Dona is among the first of its kind as an intelligent personal agent for voice assistance in student course registration. Due to its ubiquitous access for educational needs, Dona directly impacts AI for education. It makes a broader impact on smart city characteristics of smart living and smart people due to its contributions to providing benefits for new ways of living and assisting 21st century education, respectively. 
\end{abstract}

\section{Introduction}
An application called \emph{Dona} or \emph{Hey Dona} is motivated by the need to have a greater proliferation of intelligent personal agents for student course registration in the overall realm of AI for education. While there is a multitude of apps developed for several purposes, we find the need to build one for student course registration to provide virtual voice assistance, such that it caters to a variety of needs. We thus address our motivation, problem definition, related work and main contributions, with respect to our deployed project on the Dona intelligent agent. 

\subsection{Motivation and Problem Definition} 
In the modern era, we have smartphones at our fingertips, yet don't always need to use fingers. We speak of a task, and it is done, e.g. ``Text Dad: I’ll be late today'' with the text sent by a virtual voice assistant \cite{VVA}. Voice is used for rapid search, $\sim$4 times faster than written search \cite{Backlinko} (humans can write $\sim$40 words/min, speak $\sim$150). IoT devices, e.g. smart thermostats \& speakers give voice assistants high utility. Experts predict almost every application being integrated with voice technology in 5 years \cite{subhash2020artificial}. Many virtual assistants exist but may have problems. They may fathom English phrases, but fail to recognize various accents. Some are developed for mobile devices, though users would also want them on desktops for good human-AI interaction. It is desirable to get some language translation too. A virtual assistant may not able to answer all queries well, as it lacks context / misinterprets questions. In order to cater to various users and work seamlessly, it needs rigorous optimization via human interaction, machine learning and natural language instruction \cite{azaria2016instructable}. Quality control strategies can help combat issues such as background noise, e.g. in outdoor settings. Complex task dependencies should be modeled to recommend optimized plans for users. 
All these issues are significant in student course registration; hence motivating our work. We thus define our problem as the need to develop an intelligent personal agent for virtual voice assistance in student course registration, catering to issues such as different accents, indoor / outdoor settings, language translation, and desktops usage. 

\begin{figure}
    \centering
    {\frame{\includegraphics[width=1.0\columnwidth]{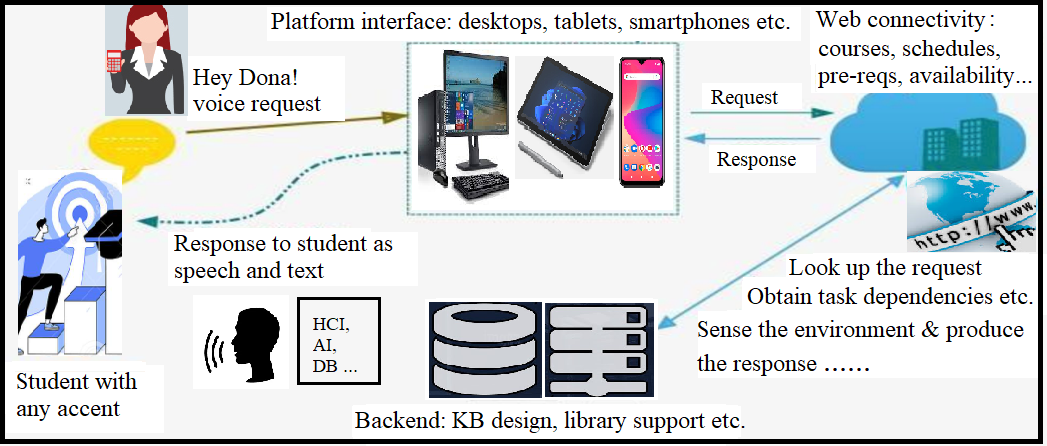}}}
    \caption{Overview of Hey Dona: Intelligent Personal Agent}
    \label{Dona}
\end{figure}

\subsection{Related Work} 
The $1^{st}$ speech recognition system, Audrey at Bell Labs in 1952 fathomed 10 digits spoken by specific people \cite{pieraccini2012audrey}. IBM built the Shoebox Machine, responding to 16 spoken words (digits $0 \dots 9$, commands $+$, $-$) \cite{rebman2003speech}. Hidden Markov Models improved speech recognition by statistical methods measuring probability of unknown sounds as words \cite{RabinerHMM}. The $1^{st}$ mass-accessible voice command system was launched by Apple as Siri in 2011 embodying NLP to answer questions and outsource requests to web services \cite{bostic2013nuance}. Hal was made by Zabaware as a virtual assistant to converse with users and take notes \cite{Hal}. IBM created Watson to compete on the TV show Jeopardy!  defeating top human players \cite{magazine2010ai}. Technologies are patented in this area, e.g. ``intelligent virtual assistant'' \cite{gong2003san}, ``Natural-language voice-activated personal assistant'' \cite{tsiao2007natural}. Though voice assistants abound, they are not too prevalent in course registration, paving the way for Dona. 

\subsection{Contributions} 
This paper presents the \emph{Dona} agent, an application deployed in AI for education, to provide a solution to the problem of virtual voice assistance in student course registration, catering to several modern-day needs. Dona or ''Hey Dona'' entails the following aspects.  
\begin{itemize}
  \setlength{\itemsep}{0pt}
  \setlength{\parskip}{0pt}
    \item Comprehend English well in various accents
    \item Work seamlessly on desktops and mobile devices
    \item Imbibe quality control, e.g. combating noise  
    \item Model task dependencies and find optimal solutions
    \item Offer translation as needed, sensing the environment
\end{itemize}
An overview of Dona appears in Figure \ref{Dona}. To the best of our knowledge, Dona is among the first personal agents for voice assistance in course registration. It impacts human-AI interaction, speech recognition, NLP and app development, in line with other work including some of our own, e.g. \cite{Conti2022}, \cite{kamath2019deep}, \cite{Corallo2022}, \cite{Razniewski2021}, \cite{li2020robutrans}, \cite{Bhagat2019}, \cite{Du2019}, \cite{Basavaraju2016}, \cite{Emebo2020}.

\section{Hey Dona: System Framework} 
The overall system framework of the intelligent agent Dona is described here. We mention its system requirements, explain its system architecture, and outline its system analysis. 

\subsection{System Requirements} 
The following hardware and software have been used in order to develop the Dona application.
\begin{itemize}
    \item {\it \bf Hardware:} A Pentium-pro processor or later is used with a RAM capacity of 512MB or more.
    \item {\it \bf Software:} We use Windows 7(32-bit)+, Python 2.7+, Chrome Driver, Selenium Web Automation, and SQLite.
\end{itemize}

\subsection{System Architecture} 
Dona is created in Python as open-source software modules with PyCharm community backing. Its main modules are described in a nutshell here. 
\begin{enumerate}
    \item {\it \bf Speech recognition:} This occurs online and offline with Google voice API. Extensive training via machine learning enables recognition of different accents and distinguishing intended speech from noise. 
    \item {\it \bf Text-to-speech:} We deploy \emph{Pyttsx} (Python Text to Speech), a cross-platform Python wrapper with common text-to-speech engines on Mac OS X, Windows, and Linux. Its \emph{pyttsx3} module supports 2 voices: female and male (by \emph{sapi5} for windows). We choose the ``female'' voice eponymous with ``Dona''.
    \item {\it \bf KB design:} DBpedia is good for Knowledge Base design because it covers many domains, represents real community agreement, automatically evolves as Wikipedia changes, and is multilingual. It helps sense the environment for translation in user conversation, as needed. 
    \item {\it \bf Library support:} SQLite (Structured Query Language Lite) is designed for library support as it has an in-process relational database (RDB) for efficient storage and querying of small-to-medium data sets. Much of this is imbibed in Dona's cache memory. Task dependencies can be modeled in-house. This is ideal as Dona only stores university-related data (e.g. courses, pre-requisites). SQLite's library runs embedded in memory alongside the application, a reasonable subset of its books are implemented by the standard library DB driver, feasible for searching user requests. Task dependencies can be modeled using RDB paradigms as SQLite thrives on SQL. 
    \item {\it \bf Audio:} PyAudio offers Python bindings for Port Audio, the cross-platform audio input-output (I/O) library. It facilitates playing and recording audio on many platforms with seamless access on desktops, smart devices etc. 
    \item {\it \bf NLP:} We harness the NLTK (Natural Language Tool Kit) of Python. It has easy-to-use interfaces for 50+ corpora and lexical resources, e.g. WordNet, with text processing suites for classification, tokenization, stemming, tagging, parsing, semantic reasoning, and an active discussion forum. It is useful in interpretation and communication for good human-AI interaction. 
\end{enumerate}

\subsection{System Analysis and Machine Learning (ML)} 
Dona is trained with vast real data on courses, voices with different accents, partially noisy settings etc. System analysis occurs to understand the framework, especially noting where it fails, e.g. various accents, complex task dependencies. Solutions are discovered to resolve issues. 

Since there are modules, their functions and interrelation can easily be studied with failure diagnosis to propose solutions. These provide the scope for further learning via ML techniques to improve performance. Neural networks with deep learning are useful, with the potential for enhancement. 

\section{User Interaction and Evaluation}
We highlight the input and output process of Dona, along with a sample of its user interaction. We outline its system workflow and provide excerpts from its user evaluation. 

\subsection{The I/O of Dona}
The basic I/O (input-output) process of Dona is synopsized in Algorithm 1. 

\hrulefill{\bf Algorithm 1: Hey Dona I/O} \hrulefill

1. {\bf Procedure RunDona()}

2. VAR chat $\leftarrow$ Predefined chats and inputs

3. While TRUE;

\hspace{15pt}	VAR CMD $\leftarrow$ TakeCommand();

\hspace{15pt}      if CMD = = NULL	repeat step 2

\hspace{15pt}	VAR RESP $\leftarrow$  Response();

\hspace{15pt}	Output  RESP;

   End While

4. End Procedure RunDona

5. {\bf Procedure TakeCommand()}

6. 	VAR source $\leftarrow$ microphone input

7. 	VAR voice $\leftarrow$ voice heard through a source

8. 	VAR CMD  $\leftarrow$ command decoded from voice

9. 	Output $\leftarrow$  CMD

10. End Procedure TakeCommand() 

\hrulefill

\bigskip 

\begin{figure}
    \centering
    {\frame{\includegraphics[width=1.0\columnwidth]{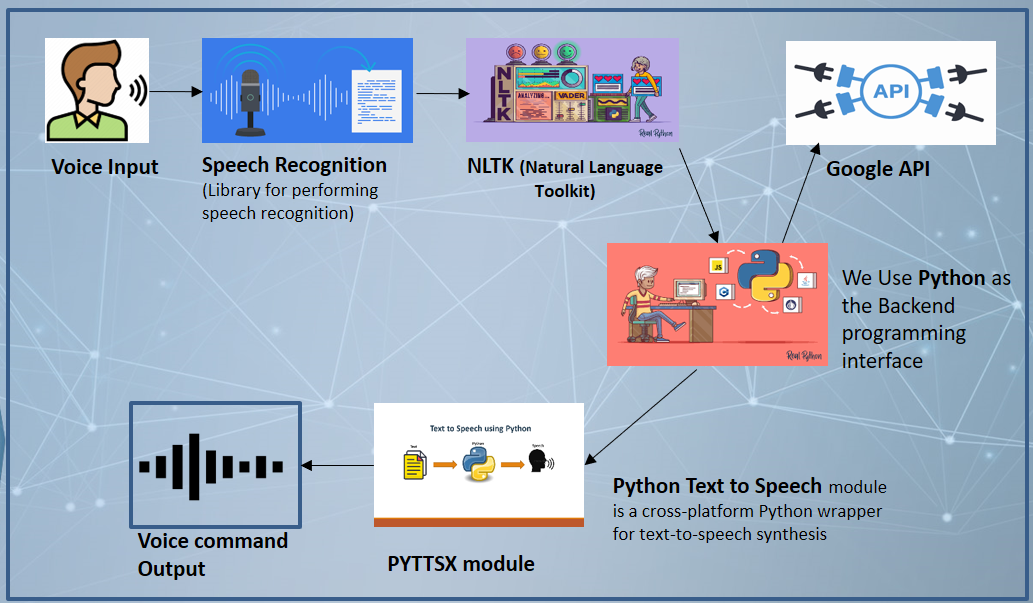}}}
    \caption{System I/O process of Dona}
    \label{IO}
\end{figure}

Figure \ref{IO} provides an illustration of the overall I/O process in Dona. 
Based on this I/O process, Listing 1 depicts a sample user interaction. \\

\noindent \hrulefill \textbf{Listing 1: User Q\&A Interaction Example} \hrulefill

\noindent Dona: How can I help you?' \\
Student: I want to register for a course. \\
Dona: What is your degree and major? \\
Student: Masters in Computer Science.\\
Dona: These courses are available... \emph{[Displays courses]} \\
Student: Register me for HCI (CSIT-535) \\
Dona: Did you complete prerequisites? \emph{[Displays them]} \\ 
Student: Yes. \\
Dona completes the registration and informs the student. 

\hrulefill
\bigskip

\begin{figure}
    \centering
    {\frame{\includegraphics[width=1.0\columnwidth]{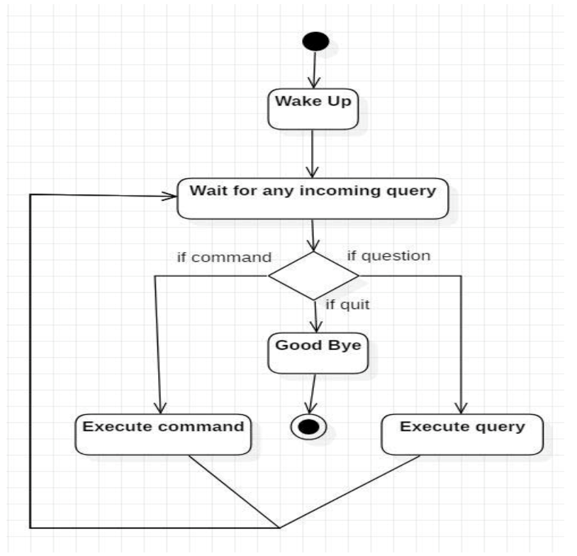}}}
    \caption{System workflow illustrating a general activity diagram for Dona}
    \label{activity}
\end{figure}

\subsection{System workflow} 
\begin{figure}
    \centering
    {\frame{\includegraphics[width=1.0\columnwidth]{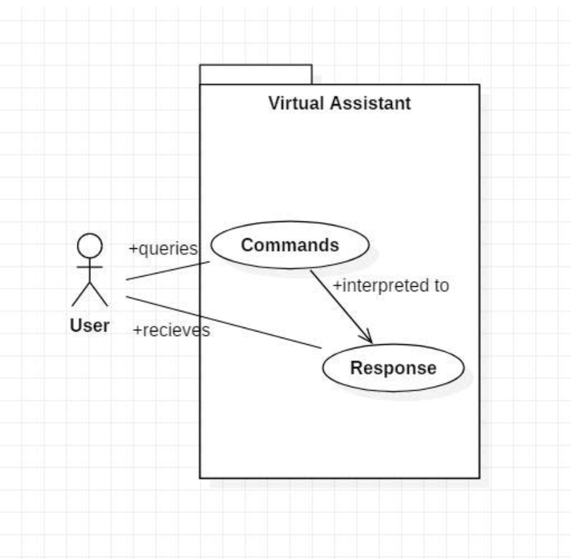}}}
    \caption{System workflow depicting a typical use case diagram for Dona}
    \label{usecase}
\end{figure}

The general system workflow in the Dona application as an intelligent agent is synopsized with an activity diagram and a use case diagram. The activity diagram for Dona's user interaction appears in Figure \ref{activity} while its use case diagram is shown in Figure \ref{usecase}. Initially, the system is in idle mode. As it receives a wake-up call, it begins execution. The received command is identified (question / task). Specific action is taken accordingly. After the question is answered or task performed, the system waits for another command. This loop continues unless it receives the quit command. At that moment, it goes back to sleep. 

On a further note as per a few implementation details in the system workflow, the pilot version of Dona demonstrated here has much of its database functionality coded into the program itself rather than being offline (to speed up the process), this being seamlessly extendable in ongoing work. Based on that, a typical user Q\&A would involve information retrieval from the MySQL database as pertinent to the user's requirements (e.g. pre-requisites, semester course offerings) via a SQL query, converting that information (query response) to text via the NLP functionality in an interactive manner, transforming that response to audio using the text-to-speech functionality through Pyttsx and communicating the voice response by deploying PyAudio, as the audio output to the user. Note that multiple conversations from the same user can be tracked because Dona asks users during a typical Q\&A interaction whether they wish to add more courses, perform any further activities etc. This is displayed in a live demo video \cite{DonaDemo}.

\subsection{User evaluation} 
Dona is assessed by 10+ student users hailing from the continents of Europe, Asia, Africa, North America and South America, and spanning countries such as USA, Netherlands, India, China, Nigeria, Brazil etc. These students have various accents and speeds of talking. Some of them are in indoor settings while some in outdoor settings. They come from different majors in their degree education and have different task dependencies based on course availability in a given semester, pre-requisite courses, and so on. They test Dona and convey their responses, which serve as the results of user evaluation. 

The evaluation results are summarized in Table 1 with round figures. In sum, all users indicate that they find the system useful. Some of them provide feedback with suggestions for improvement, e.g. some users encounter latency in Dona's processing, especially in outdoor settings, while others indicate the need for automated tuition payments to be included in Dona to make course registration seem like a more realistic experience and be truly complete. Such feedback can be incorporated in future work. A link for Dona is available on GitHub \& can be provided to interested users upon request. A video of a sample demo is included here \cite{DonaDemo} which depicts Dona's user interaction with multiple international students from various educational backgrounds. 

\begin{table}[!t]
    \centering
    \begin{tabular}{|c|c|} \hline
    {\bf Dona's Criteria}     & {\bf Students' Observation} \\ 
    \hline
    Accent Support & 90\% of users satisfied \\ \hline
    Language Translation & 75\% of users satisfied \\ \hline
    Task Optimization & 90\% of users satisfied \\ \hline
    Overall Quality & 80\% of users satisfied \\ \hline
    Problems & some latency, no tuition payment \\ \hline
    \end{tabular}
    \caption{Assessment by Various Student Users}
    \label{tab:results}
\end{table}

\section{Conclusions and Future Work}
In this paper, we present an example of a project deployed to contribute our two cents to the general realm of AI in education. This is a developed application called \emph{Dona} that offers virtual voice assistance in student course registration. User testing of Dona reveals that it is beneficial as an intelligent personal agent with human-AI interaction, NLP and speech recognition. It recognizes accents, does translation as needed, models task dependencies, works well on desktops (as opposed to some apps designed only for mobile devices), and tunes in with indoor as well as outdoor settings (e.g. in combating noise). Dona is built with PyCharm community backing to feasibly accommodate updates. 

As future work, we aim to address the following. We will to fix latency issues encountered by a few users. We intend to incorporate features to automate paying tuition bills to enrich user experiences. Furthermore, we would potentially alert students for graduation, analogous to a human advisor or a registrar office. 
Analogous to prior work in our research group \cite{Puri2018}, \cite{Amal2011}, \cite{Persaud2017}, \cite{Varghese2021}, our work in this paper relates to smart cities. It makes impacts on the \emph{smart living} \& \emph{smart people} characteristics due to its ubiquitous AI-based assistance, contributing to lifestyle quality \& $21^{st}$ century education respectively. 


\section*{Acknowledgments}
The authors thank the students who performed user evaluations of Dona, especially Isabele Bittencourt, Darko Radakovic, Iyanuolowa Shode and Xu Du, all from Montclair State University, since they provided detailed video recordings of their demos. Aparna Varde acknowledges NSF grants 2117308 (MRI: Acquisition of a Multimodal Collaborative Robot System (MCROS) to Support Cross-Disciplinary Human-Centered Research and Education at Montclair State University), and 2018575 (MRI: Acquisition of a High-Performance GPU Cluster for Research and Education). She is a visiting researcher at the Max Planck Institute for Informatics in Saarbrucken, Germany, ongoing from her sabbatical, conducting data science related research, spanning NLP and commonsense knowledge.  

\bibliography{references}
\bibliographystyle{aaai}

\end{document}